# Covid-19: Data analysis of the Lombardy region and the provinces of Bergamo and Brescia

Marco Picariello, PhD[1]; Paola Aliani, PhD[2]

**Abstract:**

The data analysis on deaths in the Lombardy Region and of both the provinces of Bergamo and Brescia shows a twofold aspect on the trend of the epidemic:

- all the data show a bias linked to the event of March 10[th] (day for which the Lombardy region data is partial) and the subsequent change in the way in which positive cases and deaths are calculated;
- following the containment measures of the Prime Minister's Decree of March 11[th], the spread of the epidemic, although still exponential in nature, has a reduced multiplication coefficient.

Our analysis concludes that the situation is not yet compatible with a *plateau* trend and allows us to predict the trend in the number of deaths in the Lombardy region.

We therefore conclude that the containment measures put in place by the government on March 11[th] will allow a reduction in deaths from around 8000 to just over 6500 for March 27[th].

## The outbreak in Lombardy

Il On February 21[st] , fourteen cases were confirmed in Lombardy. In particular, in Codogno, a town in the province of Lodi, a 38-year-old man tested positive for the virus after experiencing respiratory problems. In addition to the spouse and a friend of the man, three more cases were confirmed the same day after the onset of the symptoms of pneumonia. Subsequently, in-depth checks and controls were carried out on all the people who came into contact or who were in contact in the vicinity of infected subjects.

On February 22[nd], the first coronavirus patient died in Lombardy. The next day a 68 year old woman, affected by cancer, died in Crema, becoming the third Italian victim of the virus. On February 24[th] a 84 year-old man from Bergamo, an 88 year old from Caselle Landi and two men from Castiglione D'Adda - respectively 80 and 62 years old dies. All had previous pathologies (Wikipedia, 2020).

From February 24[th], the national system began to systematically collect data (Rosini, 2020). In our analyses, day 1 corresponds to February 24[th]. In this data analysis it is important to consider that some of the methods of detection were changed during the observation period.

In particular, the number of people actually infected does not correspond to the reported data due to the fact that it is not possible to carry out tampons on all but also on the basis that the selection criteria to whom the swab is made has changed. The data is therefore not homogeneous (Nicasto & al, 2020).

It is interesting to note that while on February 27[th] the WHO asked Italy (Caccia, 2020) not to test asymptomatic subjects in order not to trigger panic, the WHO has updated its recommendations and is currently asking to modify this method in order to contain the epidemic.

The Italian health authority recognizes all deaths from patients that have had positive Covid-19 tampon as deaths due to the virus regardless of past or underlying diseases. This is not the case in other countries. The method of attributing the cause of death has not changed since the onset of the epidemic, and therefore can be considered the most reliable data for statistical analysis.

Unfortunately, compared to other data, the number of deaths shows an 8-day delay corresponding to the median time between the appearance of the first symptoms and death, as reported by the Istituto Superiore di Sanità in the Report on the characteristics of patients who died because positive for COVID-19 in Italy (ISS, 2020).

Below is a list of containment measures published by the Italian Government, in chronological order.
- The Prime Minister's Decree "*Measures concerning the contrast and containment throughout the national territory of the spread of the Coronavirus*" (Prime Minister, 4 mar) (Government, 4 mar) of March 4[th];
- The Prime Ministerial Decree "*Further implementing provisions of Legislative Decree 23 February 2020, n. 6, containing urgent measures relating to containment and management of the epidemiological emergency from COVID-19*" (Prime Minister, 8 mar) of March 8[th];
- The Prime Ministerial Decree " *Additional measures regarding the containment and management of the epidemiological emergency from COVID-19 on the whole national territory*" (Government, 11 mar) of March 11[th] .

---

[1] MIUR – IISS L. Vanvitelli – Lioni (AV) – Italy, Marco.Picariello@istruzione.it

[2] Cognizant - Belgium

## Analysis of deaths in the whole Lombard region

The data are compatible two distinct time interval sets each with their own exponential-type trend:

$$y = a \cdot b^t \qquad \text{where } t \text{ is in days} \qquad (Eq.\ 1)$$

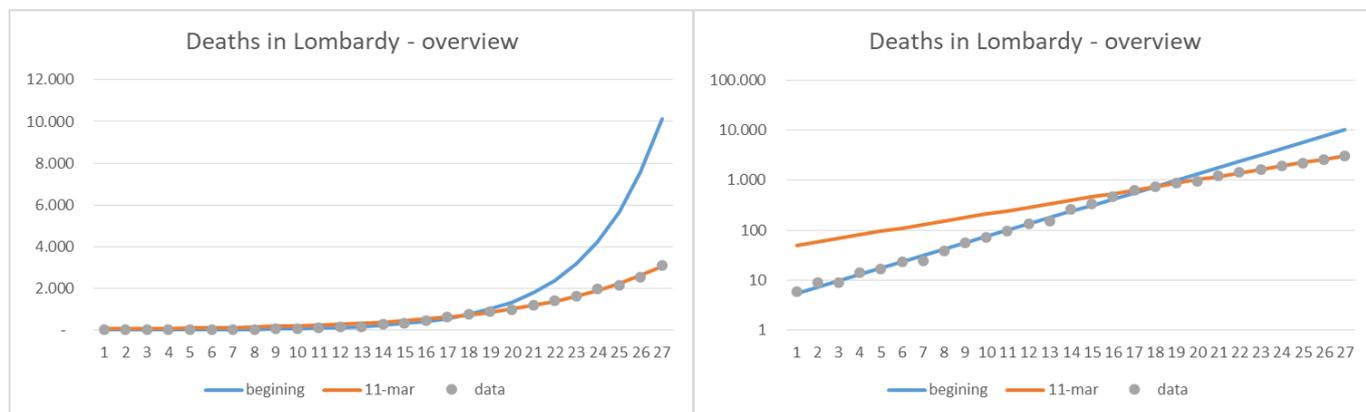

*Figure 1: Deaths in Lombardy. From the overview we can see two distinct phases in the temporal evolution.*

The presence of two phases has already been highlighted and associated (Granozio, 2020) with both a first modification of collective behaviour and individual awareness towards the end of February in the regions of the outbreak, and the first governmental initiatives.

## Initial evolution of the epidemic deaths in the whole Lombardy region

The initial exponential trend, valid for the first 16 days, has coefficients

$$b = 1{,}33 \pm 0{,}01 \qquad a = 4{,}1 \pm 0{,}1$$

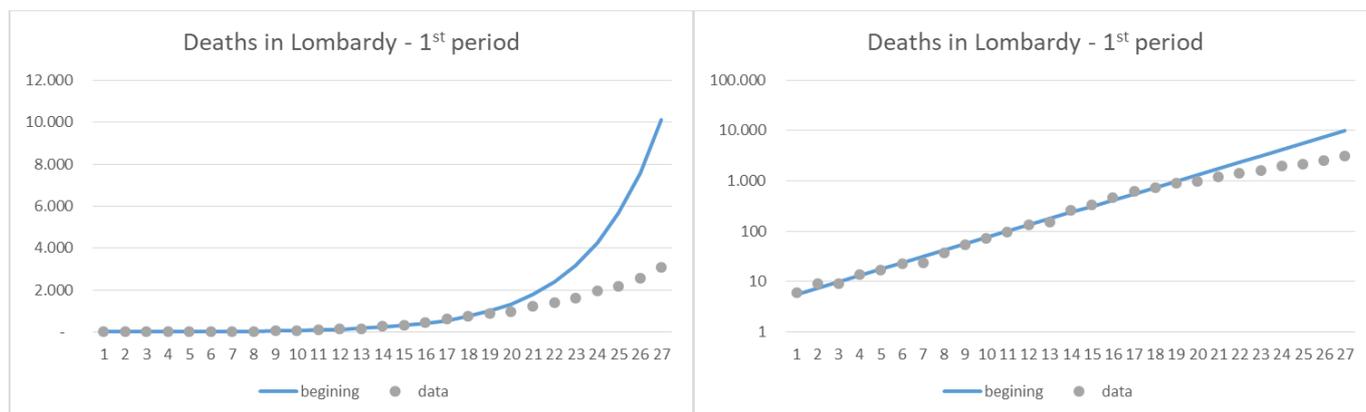

*Figure 2: Deaths in Lombardy. Initial evolution.*

## Evolution of the epidemic deaths after the 17th day throughout the Lombardy region

The second exponential trend, valid from the 17th day onwards, has coefficients

$$b = 1{,}17 \pm 0{,}01 \qquad a = 45 \pm 5$$

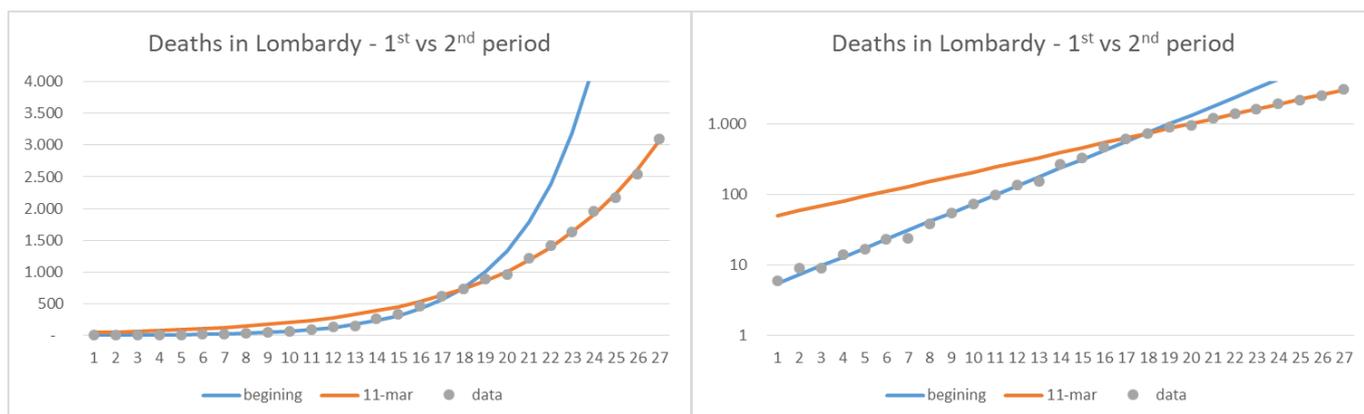

*Figure 3: Deaths in Lombardy; Transition from the 1st to the 2nd period, which took place on 11 March.*

Taking into account that from the onset of the first symptoms to death the median time is 8 days (9 if intubated) it is interesting to identify the cause of the reduction. Can we can say that the containment measures introduced on the tenth day (March 4th) with the DPCM "*measures regarding the contrast and containment on the entire national territory of the spread of Coronavirus*" (Government, 4 mar), have had the effect of reducing the speed of spread of the virus?

Up to this moment the data analysis shows no indication of any effects deriving from the DPCM nor any effects due to the saturation of the epidemic.

## An analysis of the total cases in the whole Lombardy region

In this set, the data are compatible with three time intervals each with an exponential trend, like in eq. (1).

In the data analysis we see that there seems to be a measurement error on the 16th day (March 10th). This corresponds to the day in which the Lombardy region published only partial data (Rosini, 2020). For the purpose of this analysis, the data of March 16th has been adjusted to minimise the error on the parameters involved.

As we will see, this *measurement error* does not introduce a further error in the determination of the parameters, but an uncertainty in the definition of the day of transition between the first and the second time period.

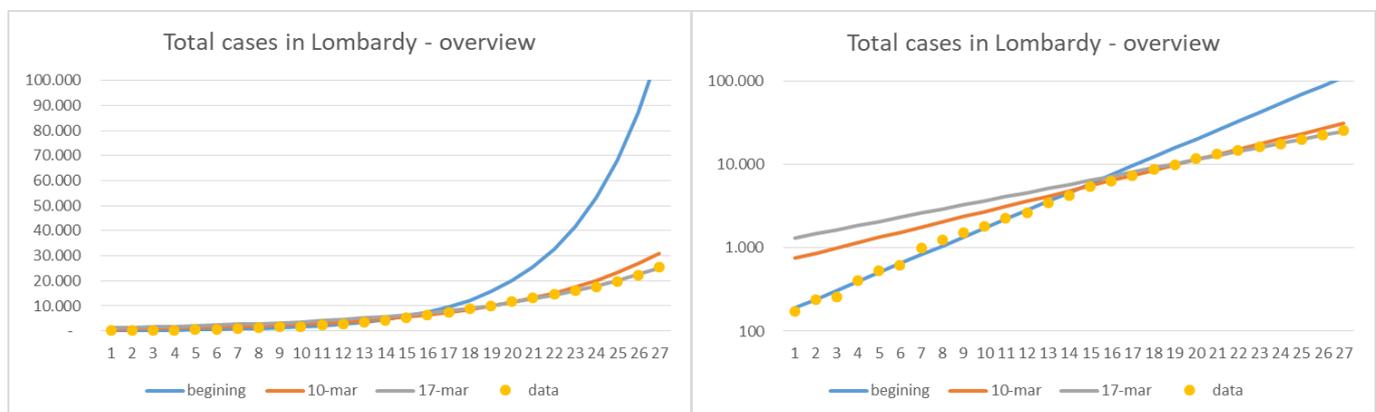

*Figure 4: Total cases in Lombardy. From the overview we highlight the three phases of temporal evolution.*

## Initial evolution of the epidemic in the whole Lombardy region

The initial exponential trend, valid for the first 15 days (until March 9th) has coefficients

$$b = 1,28 \pm 0,01 \qquad a = 150 \pm 5$$

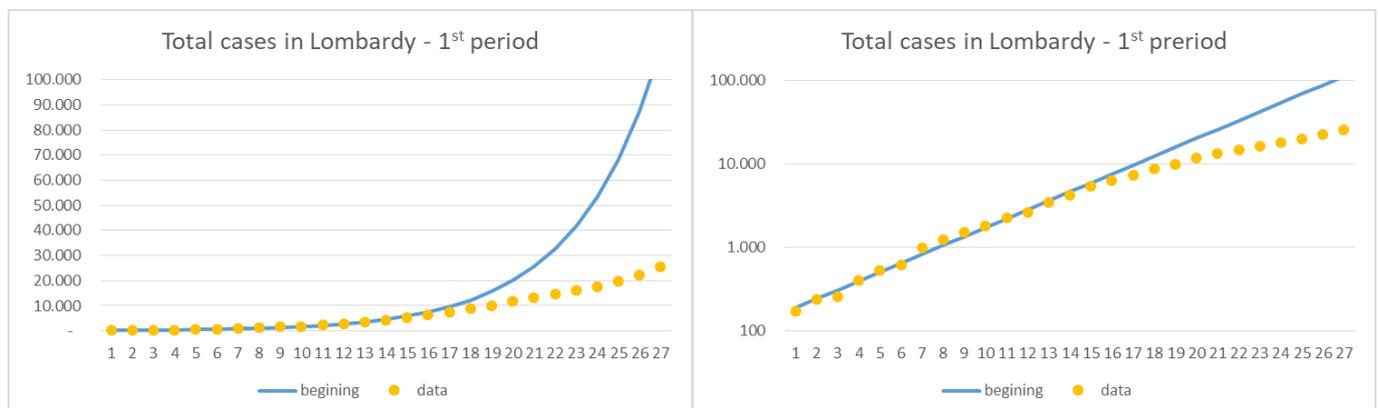

*Figure 5: Total cases in Lombardy. Initial evolution.*

As can be seen from the graph, from 10 March the data showed a slowdown in the rate of growth of the infection compared to the initial trend, and this initially made us hope that it was an arrested or whipped growth (Antonio Bianconi, 2020).

## Evolution of the epidemic after the 16th day in the whole Lombardy region

The second exponential trend, valid from the 16th day onwards (March 10th), has coefficients

$$b = 1{,}155 \pm 0{,}005 \qquad a = 650 \pm 5$$

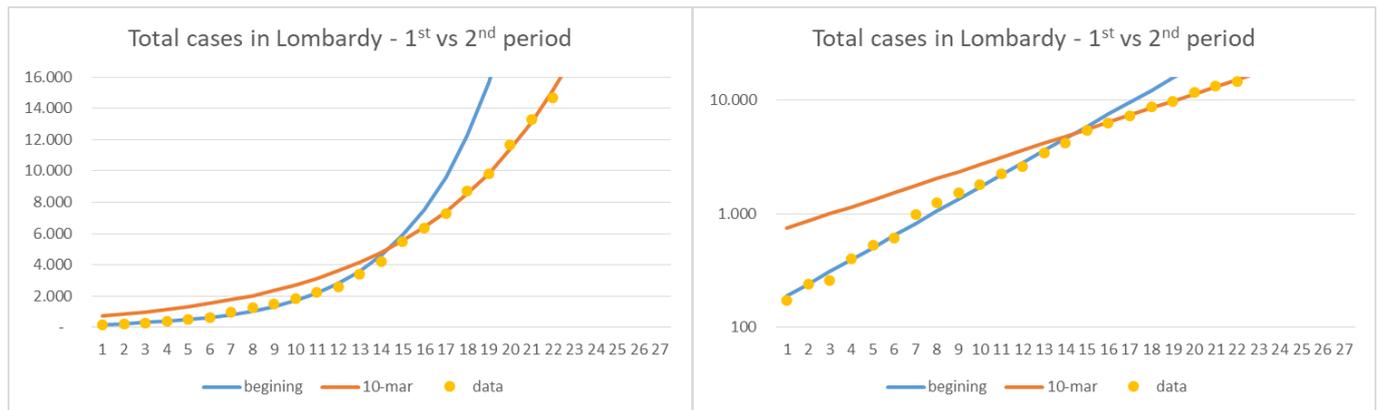

Figure 6: Total cases in Lombardy. Transition from 1st to 2nd period, which took place on March 10th.

As we have already pointed out, this change in the trend is mainly due to the different method of count of the infected and not to a real containment of the epidemic. From the data analysis it is clear that the March 10 data is not a measurement error but data completely in line with the subsequent data and therefore the *step* between 9 and 10 March highlighted by the Italian media (Wired, 2020), is nothing more than the passage from the initial evolution of the 1st period to the evolution of the 2nd period.

## Evolution of the epidemic after the 23th day in the whole Lombardy region

The third exponential trend, valid from the 23rd day onwards (March 17th) has coefficients

$$b = 1{,}12 \pm 0{,}01 \qquad a = 1175 \pm 5$$

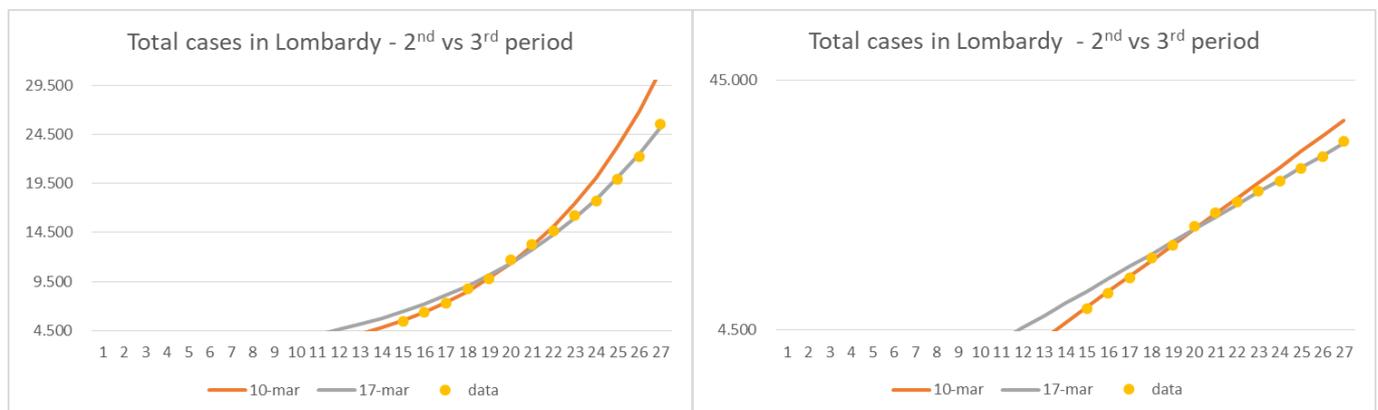

Figure 7: Total cases in Lombardy. The data show a transition phase from the 2nd to the 3rd period that took place on March 17th.

This change in performance may perhaps be related to the containment measures adopted by the government on March 11th since the time difference is compatible with the incubation period of the virus and provisioning for an average of 5 days extra, time needed for the population to perceive the containment measures and to act on them (Stephen, et al., 2020). In the *Analysis of epidemiological data of the coronavirus in Italy on March 16th* (Sebastiani, 2020) a different trend between data prior to March 19th and subsequent data had already been predicted.

## An analysis of the total cases in the province of Bergamo

The change in the trend of total cases in Lombardy could be caused by a change in the counts, which in turn can be due to the saturation and collapse of the Healthcare System in the most affected provinces. For this reason in this section we will analyse further the situation in the province of Bergamo.

The cases reported for the province of Bergamo at the beginning of the epidemic are difficult to interpret statistically because they do not follow an exponential curve. The multiple causes (containment of the epidemic, erroneous detection of cases, etc) do not allow us to analyse these initial data. For the province of Bergamo all interpolations are therefore based on data from March 1st onwards.

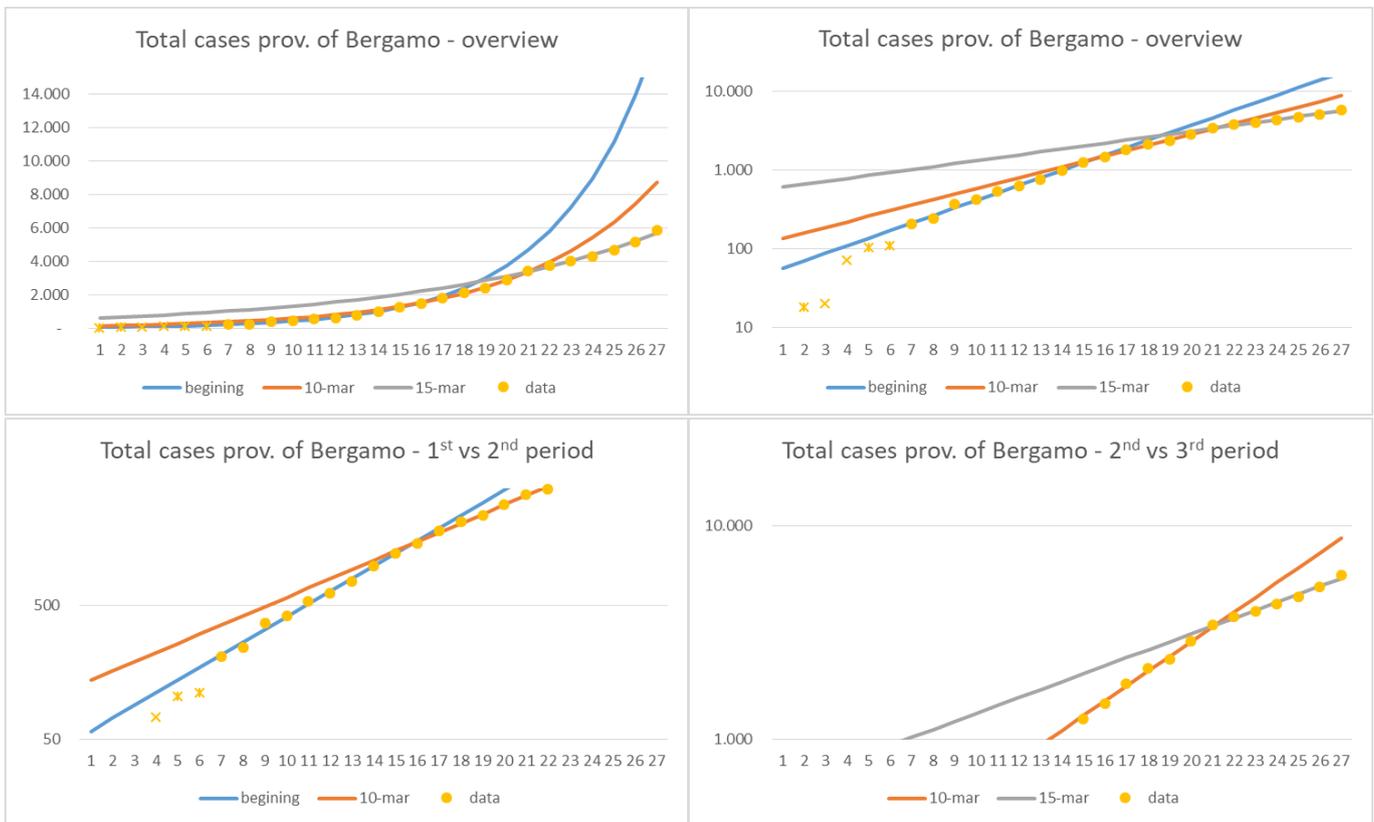

*Figure 8: Total cases in the province of Bergamo. Also in this dataset there are three time trends.*

We explicitly see that even in this set, the data are compatible with three time intervals each with an exponential trend of eq. (1).

We note, however, that while the first change takes place on March 10 (further reinforcing the hypothesis that it is an effect due to the change in the case counts), the second occurs slightly earlier than for the total Lombardy region analysis (March 15th, compared to March 17th). The three curves have the following parameters, here compared with the data of the entire Lombardy region:

|  | Lombardy region | Province of Bergamo |
|---|---|---|
| Initial parameters (from March 1st for the province of Bergamo) | $b = 1,28 \pm 0,01$ | $b = 1,25 \pm 0,02$ |
| Parameters from March 10th | $b = 1,155 \pm 0,005$ | $b = 1,17 \pm 0,01$ |
| Parameters in the third phase (begins March 15th for the province of Bergamo, March 17th for the Lombardy Region) | $b = 1,12 \pm 0,01$ | $b = 1,09 \pm 0,01$ |

It remains to be seen whether the third trend is due to the collapse of the Bergamasco Health System, and therefore to an error in the measurement of contagions, or to containment actions that will have been implemented earlier by the population in the province of Bergamo, where the seriousness of the situation was evident to all the locals.

## Analysis of the total cases in the province of Brescia

Given that due to the change of counding method that occurred on March 10 there was a modification in the trend of total cases, there remains to be determined whether the third phase is due to a change in the real trend of the epidemic or a signal of the saturation of the health system.

We have seen that in the province of Bergamo there is experimental evidence that the change occurred two days earlier than the entire Lombardy region. It is interesting to analyse whether this is also reproduced in Brescia, the second most affected province of Lombardy

As for the province of Bergamo, at the beginning of the epidemic, the reported cases are difficult to interpret statistically, since they do not follow an exponential trend and therefore we will exclude these initial data from our analysis: all interpolations are based on data from March 1 onwards.

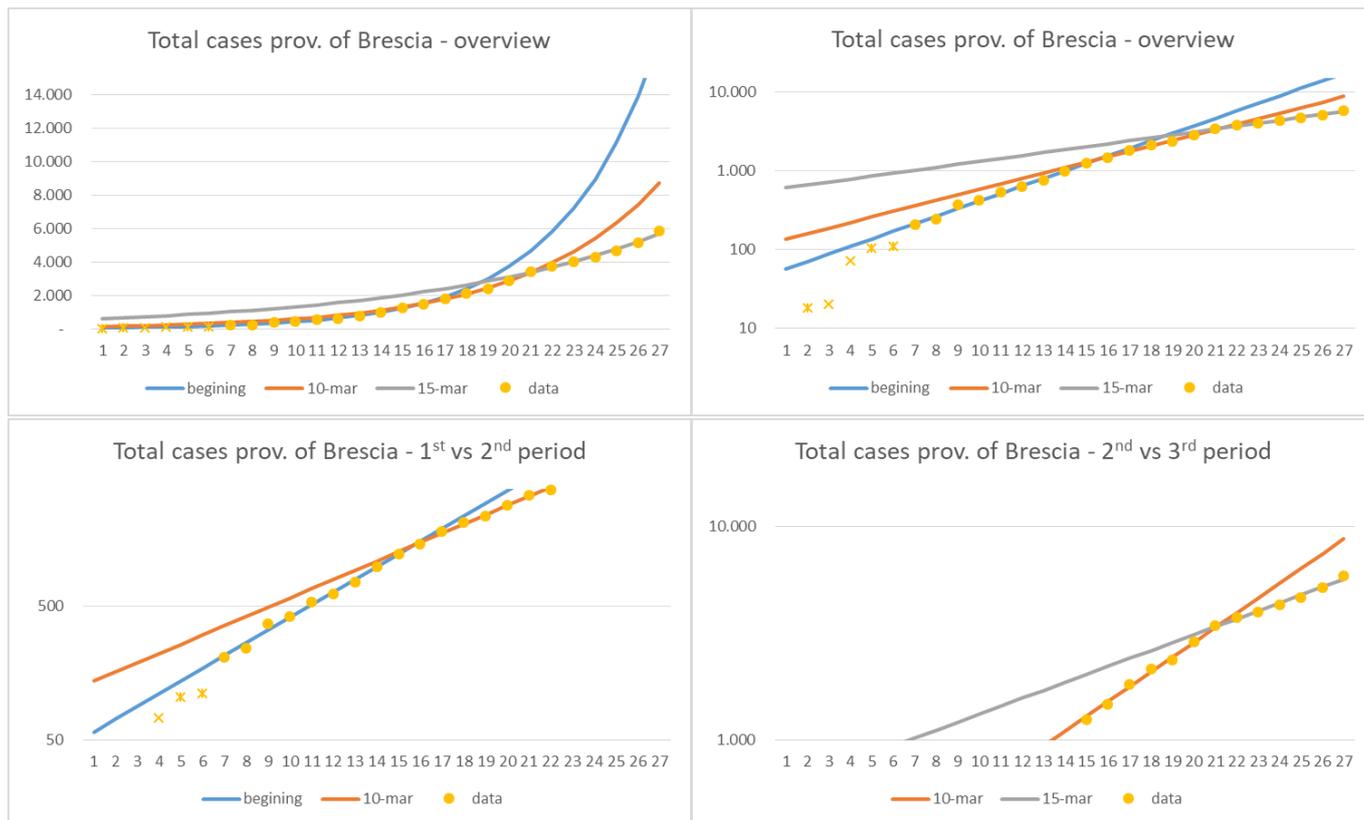

*Figure 9: Total cases in the province of Brescia. Also in this dataset there are three time trends.*

We explicitly see that even in this set, the data are compatible with three time intervals each with a exponential trend of eq. (1).

We note that, just as the first change takes place on March 10 (further reinforcing the hypothesis that it is an effect due to the change in the case counts), the second occurs temporally aligned with the data of the entire Lombardy region (March 17th).

|  | Lombardy region | Province of Brescia | Province of Bergamo |
|---|---|---|---|
| Initial parameters (from March 1st for the province of Bergamo) | $b = 1{,}28 \pm 0{,}01$ | $b = 1{,}41 \pm 0{,}02$ | $b = 1{,}25 \pm 0{,}02$ |
| Parameters from March 10th | $b = 1{,}155 \pm 0{,}005$ | $b = 1{,}21 \pm 0{,}02$ | $b = 1{,}17 \pm 0{,}01$ |
| Parameters in the third phase (starts March 15th for the province of Bergamo, March 17th for the province of Brescia and the Lombardy Region) | $b = 1{,}12 \pm 0{,}01$ | $b = 1{,}11 \pm 0{,}01$ | $b = 1{,}09 \pm 0{,}01$ |

From the point of view of temporal evolution, the province of Brescia has seen an initial explosion higher than that occurred in the province of Bergamo or in the whole of Lombardy.

Subsequently the parameters of the three curves (that of the Lombardy Region, that of the Province of Brescia and that of the Province of Bergamo) have approached and the spread of the epidemic has occurred almost homogeneously throughout the Lombard territory.

In the current phase, which began a few days in advance in the Province of Bergamo, the three curves have substantially the same parameter and therefore the spread of the epidemic is homogeneous throughout the Lombardy region.

## Short-term predictions

The first short-term prediction is that of the number of deaths in the Lombardy region. As we see from following table, even deaths in the Lombardy region follow the same trend as the epidemic, in particular the step of March 10th-11th is not affected by the delay due to the time difference from the identification of the first symptoms on the day of death.

|  | Total cases Lombardy Region | Total cases province of Brescia | Total cases province of Bergamo | Deaths Lombardy Region |
|---|---|---|---|---|
| Initial parameters (from March 1st for the province of Bergamo) | $b = 1{,}28 \pm 0{,}01$ | $b = 1{,}41 \pm 0{,}02$ | $b = 1{,}25 \pm 0{,}02$ | $b = 1{,}33 \pm 0{,}01$ |
| Parameters from March 10th | $b = 1{,}155 \pm 0{,}005$ | $b = 1{,}21 \pm 0{,}02$ | $b = 1{,}17 \pm 0{,}01$ | $b = 1{,}17 \pm 0{,}01$ |
| Parameters in the third phase (starts March 15th for the province of Bergamo, March 17th for the province of Brescia and the Lombardy Region) | $b = 1{,}12 \pm 0{,}01$ | $b = 1{,}11 \pm 0{,}01$ | $b = 1{,}09 \pm 0{,}01$ |  |

We therefore expect a trend with a coefficient equivalent to that of the total cases (b = 1.11) with a delay of approximately 8 days (median time from first symptoms to death). So until March 23 the trend of deaths should follow the current curve.

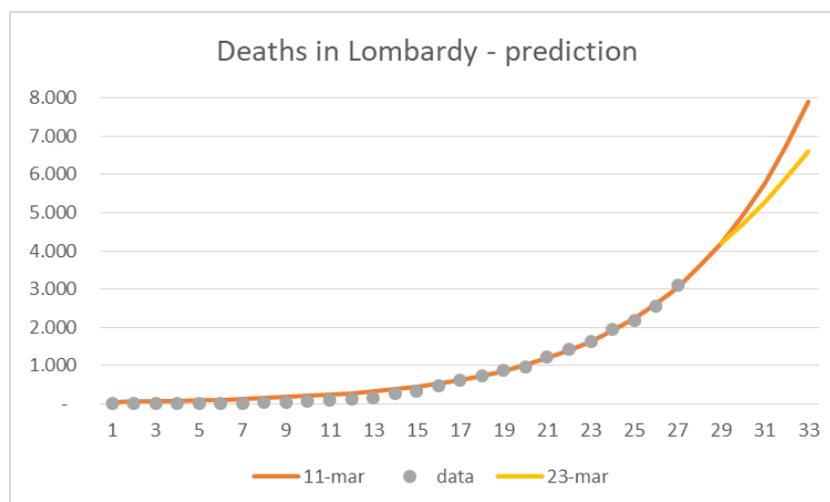

*Figure 10: prediction for the trend of deaths*

## Conclusions

We have seen how the evolution of the epidemic had a constant coefficient (excluding the modification of the counting method) until the introduction of containment measures valid throughout the territory of the Lombardy Region since March 11th. We therefore have all the evidence to conclude that the reduction in the exponential parameter of the spread of the virus is linked to containment actions. It remains to be analysed whether this reduction represents a mere modification of the exponential coefficient or if it is finally an indication that the curve is approximating the logistics expected for epidemics in general. In any case, our analysis allows us to conclude that the situation is not yet compatible with a *plateau* pattern.

To verify this conclusion, we calculated the prediction of deaths due to Covid-19 in the Lombardy Region.

Considering that the analysis performed supposes that the exponential coefficient will change starting from March 23rd, we can say that the containment measures put in place by the government on March 11th will allow a reduction in the numbers of deaths from around 8000, which would have occurred in the absence of such measures to a number of about 6500 deaths on March 27th.

It is predicted that this trend will have a change in the exponential coefficient on March 23th . Of course this prediction does not take into account the saturation of the Lombard Healthcare System, in particular of the beds in intensive and sub-intensive care.

## Acknowledgments

M.P. thanks Dr. Carla Iarrobino for the stimulating discussions and Dr. Marco Bianchetti for having placed the problem to his attention.